\documentclass[preprint,authoryear]{elsarticle-preprint}

\pdfoutput=1

\journal{Journal of Artificial Intelligence in Medicine}

\usepackage{graphicx,url,xcolor}
\usepackage{multirow,rotating}
\usepackage{amsmath,amssymb,bm}

\usepackage{booktabs}
\usepackage{wasysym}

\newcommand{\R}{\mathbb{R}}
\newcommand{\compresslist}{%
  \setlength{\itemsep}{1pt}%
  \setlength{\parskip}{0pt}%
  \setlength{\parsep}{0pt}%
}

\begin{document}

\begin{frontmatter}

\title{Leveraging Implicit Expert Knowledge for Non-Circular Machine Learning in Sepsis Prediction}

\author[icladdress,iwraddress]{Shigehiko Schamoni}
\ead{schamoni@cl.uni-heidelberg.de}

\author[ummaddress]{Holger A. Lindner}
\ead{Holger.Lindner@medma.uni-heidelberg.de}

\author[ummaddress,chsaddress]{Verena Schneider-Lindner}
\ead{Verena.Schneider-Lindner@medma.uni-heidelberg.de}

\author[ummaddress]{Manfred Thiel}
\ead{Manfred.Thiel@medma.uni-heidelberg.de}

\author[icladdress,iwraddress]{Stefan Riezler\corref{mycorrespondingauthor}}
\cortext[mycorrespondingauthor]{Corresponding author}
\ead{riezler@cl.uni-heidelberg.de}

\address[icladdress]{Department of Computational Linguistics, Heidelberg University, Germany}
\address[ummaddress]{Department of Anaesthesiology and Surgical Intensive Care Medicine, Medical Faculty Mannheim, Heidelberg University, Germany}
\address[chsaddress]{Department of Community Health Sciences, University of Manitoba, Winnipeg, Canada}
\address[iwraddress]{Interdisciplinary Center for Scientific Computing, Heidelberg University, Germany}

\begin{abstract}
  Sepsis is the leading cause of death in non-coronary intensive care units. Moreover, a delay of antibiotic treatment of patients with severe sepsis by only few hours is associated with increased mortality. This insight makes accurate models for early prediction of sepsis a key task in machine learning for healthcare. Previous approaches have achieved high AUROC by learning from electronic health records where sepsis labels were defined automatically following established clinical criteria. We argue that the practice of incorporating the clinical criteria that are used to automatically define ground truth sepsis labels as features of severity scoring models is inherently circular and compromises the validity of the proposed approaches. We propose to create an independent ground truth for sepsis research by exploiting implicit knowledge of clinical practitioners via an electronic questionnaire which records attending physicians' daily judgements of patients' sepsis status. We show that despite its small size, our dataset allows to achieve state-of-the-art AUROC scores.
  An inspection of learned weights for standardized features of the linear model lets us infer potentially surprising feature contributions and allows to interpret seemingly counterintuitive findings.
\end{abstract}


\end{frontmatter}

\section{Introduction}
\label{sec:intro}
Sepsis is the leading cause of death in non-coronary intensive care units (ICUs). Since the first consensus definition by \cite{BoneETAL:92}, also referred to as the Sepsis-1 definition, sepsis is considered a disease spectrum of increasing severity, from a systemic inflammatory response  syndrome (SIRS) to infection (sepsis), over sepsis associated with organ dysfunction (severe sepsis) to severe sepsis with hypotension despite adequate fluid resuscitation (septic shock). 
In the following years, medical research and clinical experience further advanced the guidelines for clinical management of sepsis by including variables that support the notion of sepsis in a patient, e.g. inflammatory, hemodynamic, organ dysfunction, and tissue perfusion variables. These improved guidelines are nowadays referred to as Sepsis-2 \citep{LevyETAL:03,DellingerETAL:13}. 
The third international consensus definition for sepsis (Sepsis-3) defines sepsis as a life-threatening organ dysfunction by a dysregulated host response to infection \citep{SingerETAL:16,SeymourETAL:16}. 

Despite these efforts to better define sepsis, accurate diagnosis of sepsis remains a major clinical challenge. At the same time, delay of antibiotic treatment initiation in patients at stake to develop severe sepsis and septic shock by only few hours is associated with increased mortality \citep{FerrerETAL:14,WhilesETAL:17,PeltanETAL:17}. Considerable improvement of sepsis treatment thus lies in expediting its diagnosis, i.e., by designing reliable tools for early identification of patients developing sepsis.
An avenue to enhance routine intensive care and to improve the quality of data for machine learning approaches to sepsis prediction is improved documentation in electronic health records (EHRs). Databases such as MIMIC-II \citep{SaeedETAL:11} consist of measurements of vital signals, clinical information, and laboratory test results taken at frequent intervals for populations of over 16,000 adult ICU patients. Machine learning approaches that diagnose various stages of severity from SIRS, sepsis, severe sepsis, to septic shock, as defined for example in the Surviving Sepsis Campaign Guideline (SSCG) \citep{DellingerETAL:13} or in the Sepsis-3 definition \citep{SingerETAL:16}, have successfully used this database to achieve high sensitivity and specificity in prediction \citep{HenryETAL:15,DyagilevSaria:16,NematiETAL:18}. However, the validity of these algorithms relies on knowledge of the exact time of sepsis diagnosis, and the exact manifestation of the severity stages of sepsis. Usually, none of these labels is routinely documented in databases of EHRs, nor is there a reporting standard to which machine learning approaches could refer to. Because of the lack of gold standard labels, machine learning approaches frequently resort to automatically creating ground truth labels by applying the SSCG criteria or the Sepsis-3 definition to the measurements in the EHR database.

One of the goals of this article is to call attention to a potential circularity problem that arises if the very same criteria that are used to define ground truth labels are incorporated directly as features 
into the machine learning model. We first examine this problem from an abstract viewpoint of philosophy of science where it can be described as an epistemological circle (Section \ref{sec:measurement}). Then we exemplify this circularity on the case of machine learning of risk scores for sepsis prediction (Section \ref{sec:circularity}), and present an alternative approach where an independent ground truth for sepsis research is created by exploiting implicit knowledge of clinical practitioners (Section \ref{sec:questionnaire}). Our approach is based on an electronic questionnaire (EQ) that is completed daily by senior intensivists at a surgical ICU unit, to date comprising data for a cohort of 620 patients (corresponding to over 700 admissions) collected over roughly a year. The attending physicians were instructed to make judgments based on their expert knowledge and not to strictly apply given guidelines. Despite implicit factors involved in the annotation process, the inter-rater reliability for pairs of raters for a binary distinction between sepsis (rated as Sepsis, Severe Sepsis, or Septic Shock) and non-sepsis (rated as SIRS or neither) was very high (Krippendorff's $\alpha =0.94$). In a next step we present machine learning experiments based on EQ annotations (Section \ref{sec:exps}). We report experimental results for training linear and non-linear models on this dataset, and show that despite its small size we can achieve AUROC scores that are comparable to approaches that have trained on significantly larger datasets. This finding is consistent with recent reports on the usefulness of machine learning from highly reliable judgements by medical doctors, e.g., for the tasks of sample selection \citep{Holzinger:16} or feature selection \citep{SanchezETAL:18}. We conclude our work by presenting case studies of sepsis prediction for terminally ill patients, and by discussing feature contributions by inspecting learned weights for the linear model. Our approach lets us infer contributions of features that would not have been discovered in approaches that presuppose given guidelines. Furthermore, since the linear model is intelligible, we can also explain counterintuitive findings such as seemingly positive influence of comorbidities.

\section{Measurement and Circularity in Philosophy of Science}
\label{sec:measurement}
Measurement is a central component of scientific methodology, and every concrete measurement can be characterized by a corresponding theory of measurement. Measurement theories themselves are characterized, amongst others, by the conditions that the function one wants to measure is disjoint from and uniquely determined by the other parts (functions, sets of objects, and hypotheses) of the theory \citep{Balzer:92,BalzerBrendel:19}. While several measurement theories of increasing complexity can be distinguished --- \cite{BalzerBrendel:19} discuss fundamental measurement by facts, model-derived measurement, and sample-based statistical measurement --- the crucial connection between circularity and measurement can best be described by looking at a classical example of fundamental measurement given by \cite{Stegmueller:79,Stegmueller:86}'s miniature theory of Archimedian Statics: Let us assume that this miniature theory is formalized by the set-theoretic predicate $AS$. The intended applications of the theory $AS$ are objects $a_1, \ldots, a_n$ that are in balance around a pivot point. The theory uses two functions that measure the distance $d$ of the objects from the pivot point, and the weight $g$. The central axiom of the theory states that the sum of the products $d(a_i) g(a_i)$ is the same for the objects on either side of the pivot point. The theory $AS$ can then be defined as follows:
\newtheorem{thm}{Theory AS (Archimedian Statics)}
\begin{thm} 
  {$x \textrm{ is an } AS \textrm{ iff there is an } A, d, g \textrm{ s.t. }$}
\begin{enumerate}\compresslist
\item $x = \left< A, d, g \right>$,
\item $A = \{a_1, \ldots, a_n\}$,
\item $d: A \rightarrow \R$,
\item $g: A \rightarrow \R$,
\item $\forall a \in A: g(a) > 0$,
\item $\sum_{i=1}^n d(a_i) g(a_i) = 0$.
\end{enumerate}
  \end{thm}
The problem of circularity --- or ``problem of theoretical terms'' in the terminology of \cite{Sneed:71} and \cite{Stegmueller:86} --- arises as follows: Suppose we observe persons sitting on a seesaw board. Suppose further that the board is in balance. Translating this observation into the set-theoretic language, we could denote by $y$ the balanced seesaw including the persons, and we would be tempted to make the empirical statement that 
\begin{equation}
y \textrm{ is an } AS.
\label{eq:aAS}
\end{equation}
In order to verify the central axiom, we need to measure distance and weight of the persons. Suppose that we have a measuring tape available to measure distance, and suppose further that our only method to measure weight is the use of beam balance scales. Let us denote by $z$ the entity consisting of the beam balance scale including the weighed item and the counterweight, then the validity of our measuring result depends on a statement
\begin{equation}
z \textrm{ is an } AS.
\label{eq:bAS}
\end{equation}
Thus, in order to check statement (\ref{eq:aAS}), we have to presuppose statement (\ref{eq:bAS}) which is of the very same form and uses the very same predicate. That means, in order to measure the weight of the persons, we have to presuppose successful applications of the theory $AS$. But in order to decide for successful applications of $AS$, we need to be able to measure the weight of the objects in such application. The presupposition of the validity of the theory $AS$ in the theory of measurement for the function $g$ makes $g$ an \emph{$AS$-theoretical term} and leads to an epistemological circle. A practical solution is as follows: In order to avoid the use of $AS$-theoretic terms such as $g$, we could discard the assumption that our weight measuring procedure uses beam balance scales. Instead we could use \emph{$AS$-non-theoretic measuring procedures} such as spring scales. The miniature theory $AS$ would no longer contain $AS$-theoretic terms and a circle can be avoided.\footnote{The discussion of theoretical terms is a wide field in philosophy of science, ranging from \cite{Sneed:71}'s Ramsey solution that in essence avoids theoretical terms by existentially quantifying over them, to model-derived measurement theories where equations involving the hypotheses of the very theory are solved for the values of the functions to be measured. Circularity is avoided by conditions such as disjointness and unique determination of the functions to be measured from the other functions, sets of objects, and hypotheses of the theory \citep{Balzer:92,BalzerBrendel:19}.}

\section{Circularity Problems in Machine Learning for Sepsis Prediction}
\label{sec:circularity}
Recent examples to predict sepsis by learning risk scoring functions have been presented by \cite{HenryETAL:15}, \cite{DyagilevSaria:16}, or \cite{NematiETAL:18}. In the terminology of the previous chapter, the ``theory'' of these approaches consists of a weighted combination of features established in the SSCG criteria \citep{DellingerETAL:13} or in the Sepsis-3 definitions \citep{SingerETAL:16,SeymourETAL:16}. These features are based on measurements of vital signals (respiratory rate, heart rate, blood pressure, etc.), laboratory test results (blood urea nitrogen, hematocrit, creatinine, etc.) and clinical information (clinical history, ICD-9 codes, etc.), which are incorporated as coefficients into a time-dependent linear regression function \citep{Cox:72} or into an ordinal regression function \citep{Joachims:02}. Prediction ---  or in the terminology of the previous chapter,  an ``empirical statement'' --- about sepsis, or about various stages of severity from SIRS, sepsis, severe sepsis, to septic shock, is done by identifying whether the learned severity score for a patient exceeds a certain threshold at any point of a patient's hospital stay, and by comparing the severity prediction to a ground truth label. Crucially, ground truth labels of SIRS, severe sepsis, or septic shock are obtained by taking measurements of the very same criteria based on vital, laboratory and clinical signals, that are incorporated as coefficients in the scoring function. For example, \cite{HenryETAL:15} and \cite{DyagilevSaria:16} identify SIRS by measuring whether at least two out of four SSCG criteria concerning heart rate ($>$ 90 BPM), temperature ($> 38^{\circ}$ or $<36^{\circ}$ Celsius), respiratory rate ($>$ 20 BPM), or white blood cell count ($>$ 12 or $<$ 4 thousands per microliter), measured in the last 2--8 hours, are met. These measurements of SIRS criteria are incorporated directly as features in the scoring function, and at the same time used to define a ground truth label. \cite{NematiETAL:18} incorporate all measurements required to identify a change in Sequential Organ Failure Assessment (SOFA) score \citep{VincentETAL:96} (creatinine level, Glasgow Coma Scale, bilirubin level, respiratory level, thrombocytes level) as features in their prediction function, and use them at the same time to define a ground truth label of sepsis according to the SOFA-based Sepsis-3 definition.

In the terminology of the previous chapter, in both cases the measuring procedure to establish the ground truth assumes the theory expressed by the weighted features of the scoring function to be valid, or even is part of this very theory. Supervised machine learning will optimize the parameters of the scoring function so as to meet the criteria defining the ground truth labels. For example, SIRS criteria in the ranges prescribed by the SSCG guidelines given above will receive high weights and yield optimal prediction of the SIRS label. Similar problems will arise for prediction of SOFA-based ground truth labels that can be easily learned by selecting high weights for SOFA-based features. On the one hand, the validity of such approaches is compromised by the lack of an independent ground truth against which predictions could be tested. On the other hand, the tendency of such approaches to confirm presupposed criteria may hinder the discovery of feature contributions that are not already determined by the known guidelines.

Fortunately, as has been shown in recent approaches to risk scoring functions by \cite{CaruanaETAL:15} or \cite{LiptonETAL:16}, the above described circularity is not inevitable, but can be avoided by making a clear distinction between features of the scoring function and criteria for defining the ground truth. The first approach incorporates clinical, laboratory and vital features as coefficients of a logistic regression function for the purpose of predicting pneumonia. The second approach employs a non-linear combination of similar features in a neural network to predict common ICD-9 labels. In both approaches, a clear distinction between the features incorporated in the risk scoring function and the measurement criteria for ground truth labels is made: The approach to pneumonia risk prediction of \cite{CaruanaETAL:15} uses the event of death of a patient as the ground truth label but does not include it as a feature in the scoring function; the diagnosing approach of \cite{LiptonETAL:16} makes sure to exclude ICD-9 labels from the features of the scoring function. In the terminology of the previous chapter, these approaches solve the circularity problem by using a measurement procedure that is clearly independent from the theory that is to be validated.

\section{Creating a Ground-Truth for Sepsis Prediction by Expert Evaluation of Sepsis Status in the ICU}
\label{sec:questionnaire}
\paragraph{Electronic Questionnaire} To address the above discussed desideratum of a reliable and detailed ground truth of sepsis status, we developed an Electronic Questionnaire (EQ) that aims to capture the daily opinion of treating intensivists on a patient's sepsis related health state. The details of this questionnaire were defined in close cooperation with the attending physicians where the goal was to capture sepsis-related information that cannot be directly extracted from EHR data. The attending physicians were instructed not to strictly follow the established guidelines, but to make judgments based on their expert knowledge.

The EQ is implemented on a tablet computer and is edited by senior intensivists for all patients admitted to the interdisciplinary surgical ICU of University Medical Centre Mannheim, a 1,352-bed tertiary care center. The ICU has 22 beds and is operated on a three-shift system. After an initial test phase of one month we started collecting real ground truth labels from medical experts on a daily basis. The EQ is completed for each patient in the ICU every day starting nominally at 1400h. 

The EQ consists of several sections that capture information on three domains: diagnosis, interventions, and outcomes. The driving motivations were to either record opinions based on expert knowledge or to capture sepsis-related aspects not directly encoded in EHRs or other available databases. 

\begin{description}
\item[Diagnosis.] The probably most important information is the current working diagnosis, where the physicians are able to select among 5 levels from ``Neither SIRS or Sepsis'' to ``Septic Shock'' (SIRS, Sepsis, Severe Sepsis, Septic Shock, neither). Another important question concerns the state of a (possible) infection, whether it's suspected or verified. If there is a locatable infection, we ask about the localization and whether the infection is suspected or verified. We collect detailed information on organ dysfunction, covering localization, status (new within the last 24h, acute for over 24h, or chronic), and type (infection-related, not infection-related, or reason unclear). 

\item[Interventions.] Information on infectious source control (excluding antibiotic treatment) is captured in the context of locatable infections. We also retrieve additional information on the patient's antibiotic treatment from an external database. 

\item[Outcomes.] We also ask the physicians about the change of the patient's health status during the last 24h, whether it has improved, stayed the same, or worsened. Further outcomes were extracted from the SAP's accounting system, where details on the length of stay and discharge conditions (e.g. death) are recorded. 
\end{description}

We use EQ data collected in the one year period following the test phase in May 2016. The earliest admission date is June 1st, 2016, and the latest discharge date is July 9th, 2017. 
We do not filter the patients by cause of hospital stay, but we included only sepsis patients who developed sepsis not earlier than on the second day after ICU admission in our study. If we had included patients who entered the ICU having sepsis, we would not have any pre-sepsis EHR data on these patients, which is crucial information for our predictive model (see Section \ref{sec:constructingpairs}). In total, this gives us 620 patients, with 420 non-sepsis and 200 sepsis patients. Table \ref{tab:mixedcohort} lists more details on the examined cohort.

\begin{table}[t]
\centering
\begin{tabular}{rccccc}
 \toprule
 & Patients &(male/female) & $\diameter$ Age & $\diameter$ ICU days \\
 \midrule
 Non-sepsis & 420 &(227/193) & 60.468 (61) & 4.41 (2) \\
 Sepsis     & 200 &(124/76) & 62.467 (63) & 23.3 (17) \\
 Total      & 620 &(351/269) & 61.123 (63) & 10.5 (5) \\
 \bottomrule
\end{tabular}
\caption{Details on the cohort we examine in our experiments. Average values ($\diameter$) are followed by the median value in parenthesis. For sepsis patients, the average number of ICU days before the first sepsis episode was 7.4 (median 5.8).}
\label{tab:mixedcohort}
\end{table}

\paragraph{Inter-rater Reliability} To assess the agreement between the three participating expert raters, we conducted an inter-rater reliability analysis on 127 patients. On six days a second intensivist was asked to edit a shortened paper version of the questionnaire. Days were selected such that each possible pair of physicians rated on two days. 

Rater judgements were given according to five categories 1=SIRS, 2=Sepsis, 3=Severe Sepsis, 4=Septic Shock, and 0=neither. In our experiments we used a binary distinction of non-sepsis labels (categories 0 or 1) versus sepsis labels (categories 2, 3, or 4).
Agreement was assessed calculating Krippendorff's $\alpha$-reliability \citep{Krippendorff:13} for pairs of raters for the binary distinction, yielding $\alpha=0.94$ ($0.86$--$1.0$ at a $95\%$ confidence interval).\footnote{Krippendorff's $\alpha$ agreement was chosen to account for missing raters in the pairwise agreement of three raters.}

\paragraph{Expert Labels versus Automatic Labeling} In order to examine our hypothesis that expert labels include information that differs and goes beyond labels that are extracted automatically from EHR data, we examined the correlation between gold expert labels and algorithmically generated Sepsis-3 labels. The Sepsis-3 labels were constructed as proposed in \citet{SeymourETAL:16} and we considered a 2 point change in SOFA as a criterion in addition to the condition of suspected infection window. To account for hourly differences of sepsis outcome between our expert and the algorithmically generated labels, we aligned all labels to days before comparison. Our computation of Cohen's $\kappa$ yielded a value of $0.34$, which is to be considered minimal or weak agreement according to modern studies on inter-rater reliability \citep{LeBretonSenter:08,McHugh:12}.\footnote{Very similar results were obtained by computing related agreement measures such as Scott's $\pi$ or Krippendorff's $\alpha$.}

\section{Sepsis Prediction and Evaluation: Models and Empirical Results}
\label{sec:exps}
\subsection{Models}
\label{sec:models}

Our learning goal is to optimize an ordinal regression objective \citep{HerbrichETAL:00} that allows to learn a correct order of different severity stages of patient's conditions. For each feature pair $(\mathbf{x}_i,\mathbf{x}_j)$ of different severity stages, let $\mathbf{x}_i$ represent a more severe condition than $\mathbf{x}_j$, 
and let $\mu_{i,j}^{\mathbf{w}}$ define a margin of separation of a pair $(\mathbf{x}_i,\mathbf{x}_j)$ with respect to a scoring function $g_{\mathbf{\theta}}(\cdot)$ as
$$ 
	\mu_{i,j}^{\mathbf{w}} = g_{\mathbf{\theta}}(\mathbf{x}_i) - g_{\mathbf{\theta}}(\mathbf{x}_j).
$$ 

We follow a maximum-margin approach where the goal is to learn the parameters $\mathbf{\theta}$ s.t. the number of correctly ordered tuples is maximized, and the minimal margin $\mu_{i,j}$ is maximized. Note that this formulation is agnostic of the scoring function $g_{\mathbf{\theta}}(\cdot)$. We conduct experiments evaluating linear functions \citep{Joachims:02}
\begin{equation}
g_{\mathbf{\theta}}(\mathbf{x}) = \mathbf{w}^\intercal\mathbf{x},
\label{eq:linearmodel}
\end{equation}
as well as non-linear functions \citep{RumelhartETAL:86} 
\begin{equation}
g_{\mathbf{\theta}}(\mathbf{x}) = \mathbf{w}^{(3)}\cdot( \tanh( \mathbf{b}^{(2)} + \mathbf{W}^{(2)}( 
	\tanh( \mathbf{b}^{(1)} + \mathbf{W}^{(1)} \mathbf{x})))),
\label{eq:nonlinearmodel}
\end{equation}
where $\mathbf{\theta}$ represents the model parameters ($\mathbf{w},\mathbf{b},\mathbf{W}$), and $\mathbf{x} = \mathbf{x}_i - \mathbf{x}_j$ is a difference vector representing a data pair. The non-linearity $\tanh$ is applied element-wise. In all cases, we train the model with stochastic gradient descent (SGD) where the error is calculated based on hinge loss. 

\subsection{Feature Sets}

Features of the model are extracted from data routinely collected in a hospital setting, e.g., EHRs extracted from a patient data management system (PDMS), accounting system records (for pre-existing conditions), or discharge records (for hospital stay duration, or demographic features like age and sex). The PDMS is combined with an hospital information system as well as available microbiology and laboratory test data. This database is made available for in-house research use after anonymization. The process is supervised by the data security board at the UMM. The ethics approval was obtained from the Medical Ethics Commission II of the Medical Faculty Mannheim, Heidelberg University (reference number 2016-800R-MA). 

As the PDMS collects data from different sources, the time lines of measurements (i.e., features) have different time resolutions. To construct a complete set of features at each time step during ICU stay, we establish a carry-forward strategy where the previous measurement is ``carried forward'' until a new value is available. This approach takes into account the clinical practice that measurements of critical features are likely to be repeated in the near future, while normal values are only checked in routine intervals. At start of admission, there are naturally some clinical measurements missing. Such unknowns are set to default values defined by an experienced clinician (see Table \ref{tab:stdfeatures}). 

We standardize all features using the $z$-transformation $z=\frac{x-\mu}{\sigma}$, where $\mu$ is the mean and $\sigma$ the standard deviation of the feature value. Features were standardized across all patients and not on a per-patient basis.

\paragraph{EHR Feature Set} 

The feature set extracted from EHRs was constructed to achieve a reasonable coverage across all patients in the cohort. We included all routinely monitored and stored clinical parameters, blood gas analysis values and standard laboratory test results. The final feature set consists of 42 features in different time resolution extracted from the PDMS system. 
These features were checked for implausible values which were removed, i.e. falling back to the imputation mechanism described above. Plausibility intervals were defined by clinicians from the ICU. Table \ref{tab:stdfeatures} lists all features by clinical name, their default values, and valid ranges. Three features are derived features:
\begin{enumerate}
\item Horowitz index: ratio of partial pressure of oxygen in arterial blood (PaO$_2$), and the fraction of oxygen in the inhaled air (FiO$_2$).
\item BUN-to-creatinine ratio: ratio of blood urea nitrogen (BUN) and serum creatinine.
\item $\Delta$-temperature: absolute difference to the normal mean of 37$^\circ$ Celsius.
\end{enumerate}
In addition, we incorporated three demographic features that encode sex and age of a patient.

\paragraph{Pre-existing Conditions (PEC)} Based on the Charlson Comorbidity Index (CCI) we included medical conditions by their ICD-10 code as additional features in our model. \citet{CharlsonETAL:87} classify comorbid conditions which potentially alter the risk of mortality in patient studies. To connect the comorbid conditions to the ICD-10 codes available in our dataset, we followed the work of \citet{SundararajanETAL:06}. Due to our data-driven approach, we keep only the 10 most frequent ICD-10 codes, i.e. codes that appeared at least 20 times in our patient's cohort. See Table \ref{tab:weightsicd} for the list of included codes. 

\subsection{Training Data Construction}
\label{sec:constructingpairs}

To learn a model that is able to discriminate between different severity stages, we consider two types of training pairs: 
\begin{enumerate}
\item Inter patient: compare a snapshot in time of a non sepsis-labeled patient (negative examples) with a snapshot in time of a sepsis-labeled patient (positive examples). 
\item Intra patient: compare a snapshot in time of a patient preceding the first sepsis episode (negative examples) to a snapshot in time after sepsis diagnose (positive examples). 
\end{enumerate}
Both types of training pairs are applied in an alternating manner.

In order to increase predictability of the score, positive examples are actually sampled from a 24h interval preceding a sepsis diagnose, and negative examples from a period of 24h or more preceding a sepsis diagnose. This is done because (a) not restricting the time frame might give us samples from an already recovered stage or a second episode, and (b) interventions applied after sepsis onset would influence the values of our features. 

\subsection{Optimization Settings}

Our linear model follows Equation \ref{eq:linearmodel} where we learn specific weights $\mathbf{w} = (w_1,\dots,w_n) $ for each of $n$ features of the model. We learn a score that orders patients snapshots in time according to the severity of their condition, effectively learning a score that discriminates between different severity stages of patients. All linear models are trained on 800,000 training pairs for 1,000 epochs with early stopping and a learning rage of 1e-5. 

By $z$-scoring all features the weights learned by our model are comparable between different values and stay interpretable. See Section \ref{sec:conc} for a detailed discussion. In general, positive weights increase the score and negative weights decrease the score when the corresponding feature value goes up. 

The non-linear model follows Equation \ref{eq:nonlinearmodel}. In detail, the non-linear model consists of two hidden layers each of size 300. We trained the model on the same 800,000 training pairs as the linear models and for 500 epochs with a minibatch size of 200 and a learning rate of 2e-6.  Hyperparameters were tuned on a dev set extracted from the training set. In addition to early stopping, we applied dropout \citep{SrivastavaETAL:14} with a ratio of 0.5 for the first hidden layer and 0.2 for the second layer.

\subsection{Evaluation}

\begin{table}[t]
\centering
\begin{tabular}{rcccc}
 \toprule
   & \multicolumn{2}{c}{Linear model} & \multicolumn{2}{c}{Non-linear model} \\
 time & EHR & EHR+PEC & EHR & EHR+PEC \\
 \midrule
 4h      & 83.0 \scriptsize{(81.3--84.7)} & 83.3 \scriptsize{(81.8--84.8)} & 83.6 \scriptsize{(81.5--85.5)} & 83.7 \scriptsize{(81.5--85.8)}\\
 8h      & 80.1 \scriptsize{(77.9--82.4)} & 80.4 \scriptsize{(78.2--82.5)} & 80.8 \scriptsize{(78.1--83.6)} & 81.3 \scriptsize{(78.7--83.9)}\\
 12h     & 80.5 \scriptsize{(77.8--83.3)} & 80.7 \scriptsize{(78.2--83.0)} & 80.9 \scriptsize{(77.7--84.0)} & 81.5 \scriptsize{(78.5--84.4)}\\
 24h     & 64.8 \scriptsize{(60.3--69.6)} & 67.0 \scriptsize{(61.9--72.0)} & 65.2 \scriptsize{(60.2--70.3)} & 67.0 \scriptsize{(61.4--72.6)}\\
 12h--8h & 80.7 \scriptsize{(78.3--83.3)} & 80.8 \scriptsize{(78.6--83.0)} & 81.3 \scriptsize{(78.5--84.2)} & 81.7 \scriptsize{(78.9--84.4)}\\
 24h--12h& 76.2 \scriptsize{(72.9--79.3)} & 77.0 \scriptsize{(73.9--80.1)} & 76.9 \scriptsize{(73.3--80.3)} & 77.6 \scriptsize{(73.9--81.1)}\\
 \bottomrule

\end{tabular}
\caption{AUROC results of the score for linear and non-linear models predicting sepsis at different time points and intervals before sepsis onset. \emph{EHR} models were trained on EHR features only, \emph{EHR+PEC} models additionally incorporate pre-existing conditions. Ranges in parenthesis denote the 95\% confidence intervals.}
\label{tab:modelcomparison}
\end{table}

We evaluate our models by calculating the area under curve of the receiver operator characteristic (AUROC) at different points and intervals in time preceding the onset of sepsis. A sepsis patient whose score if above the threshold at time $t$ or in interval $(t,u)$ is considered a true positive. A non-sepsis patient whose score is above the threshold any time during hospital stay is a false positive, as is a sepsis-patient whose score is above the threshold before time $t$. 

\begin{figure}[t]
\centering
\begin{minipage}[t]{0.5\linewidth} 
	\includegraphics[angle=-90,width=6cm]{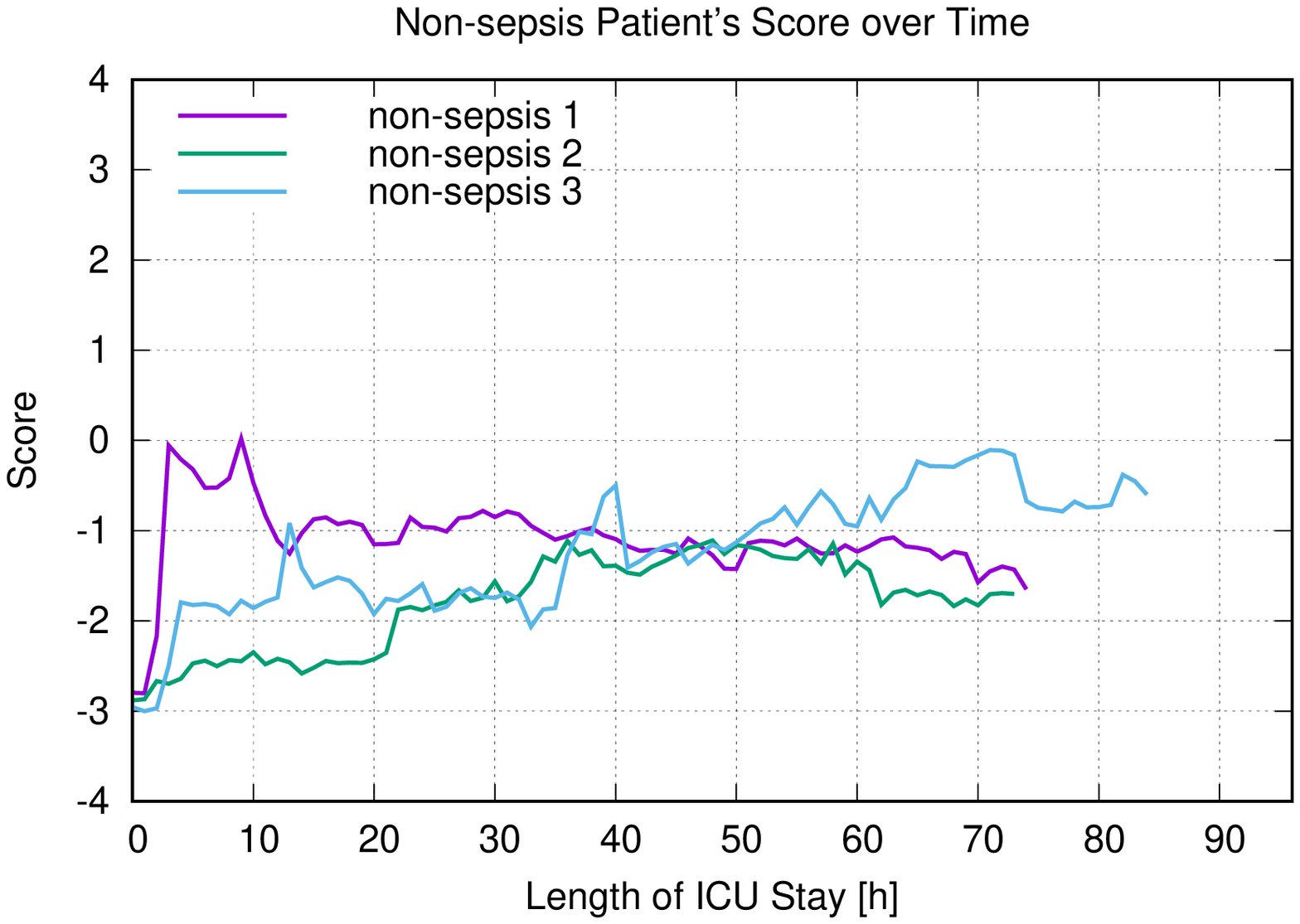}
\end{minipage}
\begin{minipage}[t]{0.49\linewidth} 
	\includegraphics[angle=-90,width=6cm]{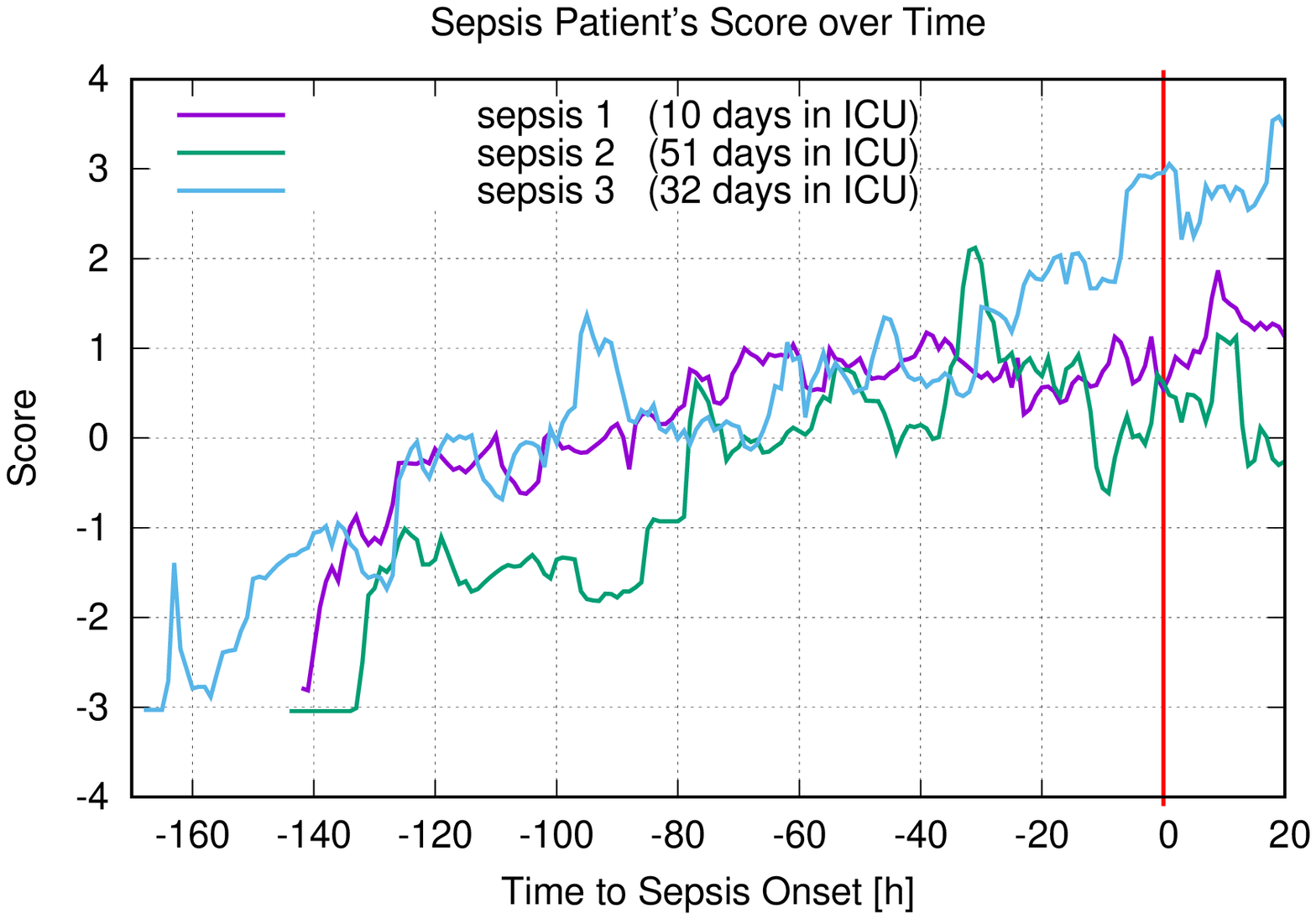}
\end{minipage} 
\caption{Example scores of non-sepsis and sepsis patients. The red vertical line in the right figure indicates the onset of the first sepsis episode. For non-sepsis patients, time lines for the length of stay in the ICU are plotted. For sepsis patients, the length of stay is added in parenthesis to the label.}
\label{fig:sepsisnonsepsisscore}
\end{figure}

More specifically, we compute the score over the whole time of stay for non-sepsis patients, and over the time range before the first sepsis episode for sepsis patients. In consideration of the different time resolution of each feature, we calculate the score in 10 minute intervals and then average over the preceding hour. We then evaluate our model at time steps 4h, 8h, 12h, and 24h, as well as time intervals 12h--8h and 24h--12h before sepsis onset for each sepsis patient, and over the whole time of stay for non-sepsis patients. Figure \ref{fig:sepsisnonsepsisscore} shows plots of the score over time for six patients, three non-sepsis (left) and three sepsis patients (right). The vertical line in red (right) indicates the time of sepsis diagnosis. In the case of non-sepsis patients, the patients were selected based on similar length of total stay, whereas in the case of sepsis patients, they were selected based on similar time to onset.

Although the pairwise ranking approach enables us to generate informative pairs in large numbers, our patient's cohort is still relatively small for a pure data-driven approach. Thus, we use 10-fold cross validation and calculate the macro-average of the achieved AUROC values. Training splits comprise 378 non-sepsis and 180 sepsis patients, and testing splits 42 and 20 patients, respectively. We used the same data splits across all experiments. Confidence intervals were calculated using bootstrapping \citep{Efron:93}.

Table \ref{tab:modelcomparison} presents AUROC results for linear and non-linear models, with and without the addition of PEC features. Best results are obtained for non-linear models including PEC features, yielding AUROC of $81.7$ at the 12h--8h interval before sepsis onset. Despite not directly comparable, this score is in the range of results reported by \cite{HenryETAL:15} (predicting septic shock at an AUROC of $83$) or \cite{NematiETAL:18} (predicting Sepsis-3 at AUROC of $83$) for a similar prediction interval.

\begin{table}[th]
\centering
\begin{tabular}{rrcrrc}
 \toprule
feature name & \multicolumn{1}{c}{weight} & \multicolumn{1}{c}{$\sigma$}
& feature name & \multicolumn{1}{c}{weight} & \multicolumn{1}{c}{$\sigma$}\\
 \midrule
C-reactive protein & 0.60482 & 0.0221 & Hemoglobin & $-0.28263$ & $0.0258$ \\
Thrombocytes & 0.47123 & 0.0318 & Stroke volume & $-0.18738$ & 0.0314 \\
Heart time volume & 0.38900 & 0.0184 & Creatinine & $-0.17967$ & 0.0322 \\
Sodium & 0.30716 & 0.0177 & Calcium & $-0.14077$ & 0.0412 \\
BUN-to-creat.~ratio & 0.26800 & 0.024 & Arterial pH & $-0.13287$ & 0.0235 \\
Chloride & 0.24971 & 0.0324 & Gender female & $-0.11281$ & 0.0264 \\
Procalcitonin & 0.22304 & 0.0372 &  Aspartate & \multirow{2}{*}{$-0.10056$} & \multirow{2}{*}{0.0301} \\
Blood urea nitrogen & 0.21703 & 0.0319 & aminotransferase & & \\
Potassium & 0.19223 & 0.0122 & O$_2$ saturation & $-0.09960$ & 0.0094 \\
Respiratory rate & 0.18556 & 0.0216 & Respiratory & \multirow{2}{*}{$-0.07849$} & \multirow{2}{*}{0.0332} \\
$\Delta$-temperature & 0.17002 & 0.0182 & minute volume & & \\
Leukocytes & 0.14518 & 0.0276 & Horovitz index & $-0.07020$ & 0.0198 \\
Heart rate & 0.12549 & 0.0239 & Diastolic BP & $-0.04308$ & 0.0149 \\
Gender male & 0.11281 & 0.0264 & Mean BP & $-0.03686$ & 0.0109 \\
Temperature & 0.10599 & 0.0241 & Quick score & $-0.03273$ & 0.0472 \\
Partial CO$_2$ & 0.10490 & 0.0111 & Systemic Vascular & \multirow{2}{*}{$-0.02782$} & \multirow{2}{*}{0.0354} \\
Pancreatic lipase & 0.10416 & 0.0429 & Resistance Index & & \\
Lactate & 0.09350 & 0.0223 & Bilirubine & $-0.02346$ & 0.0286 \\
Fraction of inspired O$_2$ & 0.08299 & 0.0242 & Urine output & $-0.01966$ & 0.0242 \\
Bicarbonate & 0.06412 & 0.0082 & & & \\
Oxygenation saturation & \multirow{2}{*}{0.05804} & \multirow{2}{*}{0.0169} & & & \\
(mixed venous) & & & & & \\
Netto (fluids) & 0.05410 & 0.0060 & & & \\
Alanine transaminase & 0.04666 & 0.0361 & & & \\
Lymphocytes & 0.04047 & 0.0195 & & & \\
Blood glucose & 0.03827 & 0.0161 & & & \\
Base excess & 0.01382 & 0.0106 & & & \\
Age & 0.01344 & 0.0319 & & & \\
Systolic BP & 0.00118 & 0.0184 \\
Part. pressure art. O$_2$ & 0.00049 & 0.0186 \\
 \bottomrule
 
\end{tabular}
\caption{Averaged feature weights sorted in descending order of absolute weight and corresponding standard deviations $\sigma$ of the linear model for 42 EHR plus 3 demographic features. For positive weights, a rising value increases risk (left half), while for negative weights a rising value decreases risk (right half). }
\label{tab:weightsfull}
\end{table}

\subsection{Interpreting Model Weights}

Linear models have the advantage of being intelligible, at a small tradeoff in accuracy compared to non-linear models (amounting to less that 1 AUROC point in our evaluation, see Table \ref{tab:modelcomparison}). Intelligibility of linear models is based on sorting features by their importance. Several measures have been proposed for this purpose, including measures based on the gradient of the learned scoring function with respect to an input feature \citep{NematiETAL:18}, to measures based on the drop in AUROC when removing a particular features \citep{CaruanaETAL:15}. We apply a simple standardization by $z$-transformation of features \citep{Schielzeth:10} that allows the magnitudes of learned weights to be compared across features. While features based on clinical measurements are almost never completely uncorrelated, a decorrelation of input features by principal component analysis is inappropriate if an interpretation of the contribution of these very features is sought. Interpretation of feature importance by comparing model weights has to keep this in mind and needs to be done with precaution.

Table \ref{tab:weightsfull} lists all weights and standard deviations $\sigma$ calculated on 10 folds for the 42 EHR features and 3 demographic features, sorted in descending order by absolute weight and grouped into positive (left) and negative weights (right). As our model predicts a severity score, increasing values of features with positive weights and decreasing values of features with negative weights both increase risk. On the other hand, decreasing values of features with positive weights as well as increasing values of features with negative weights reduce risk.

Weights close to zero with relatively high $\sigma$ are located at the end of both negative and positive groups in Table \ref{tab:weightsfull} and can be considered ambiguous in their effect on the severity score: their contribution is not clearly positive or negative. 

Highest risk features were cardiac output, thrombocytes, and C-reactive protein (CRP), in increasing order. High CRP levels are strongly indicative for infection, while increased cardiac output can have many causes: from a higher heart rate due to sheer stress, to issues in oxygen delivery. A high thrombocyte concentration as a high risk factor to develop sepsis, however, might be a potentially surprising discovery. Thrombocytopenia is often encountered in sepsis patients, and decreasing levels are associated with an increasing SOFA-score in its hematologic component. Our model assigns the second highest weight to the $z$-scored thrombocyte feature, i.e., increasing risk with higher concentration levels, contradicting the SOFA-based Sepsis-3 definition in our experiment. This is in accord with \cite{StoppelaarETAL:14} who show that high circulating platelet concentrations prior to the onset of sepsis might significantly enhance the host's inflammatory response to bacterial insults, thus aggravating microcirculatory disturbances and ensuing organ dysfunction.

On the other side of the list, highest risk features were low hemoglobin levels and low stroke volume indicating problems in oxygen transport as in hypoxia. Low calcium and creatinine levels can be indicative for kidney related dysfunction. However, the third highest negative weight for creatinine is another potentially surprising discovery. The renal SOFA-score component is defined on increasing creatinine levels, thus our learned feature weight again contradicts the Sepsis-3 definition. From a clinical point of view, however, a decrease in sepsis risk by an increase in serum creatinine might be explained again by thrombocyte dysfunction due to retention of uremic toxins \citep{BladelETAL:12}. 

Other factors like acidosis are less surprising, as this condition is common in critically ill septic patients. Our model identifies an acidic arterial pH and at the same time high lactate levels as high risk factors, which is in perfect compliance with other clinical studies \citep{RiversETAL:01}. Unquestionable, further discussion on the interpretation of identified factors for sepsis prediction with clinical experts is essential.

In addition to this, it is remarkable to see two derived values, namely Horovitz index and BUN-to-creatinine ratio, as important factors in both groups of the list. The Horovitz index measures the lung function by calculating the partial pressure of oxygen in arterial blood (PaO$_2$) and the fraction of oxygen in the inhaled air (FiO$_2$). It is also part of the respiratory component of the SOFA-score. The BUN-to-creatinine ratio determines kidney problems, as reabsorbed blood urea nitrogen (BUN) can be regulated while filtrated creatinine can not. Thus, both values indicate issues in regulation of either lung or kidney function, underlining their usefulness to identify a dysregulated host response as in sepsis. 

Another derived feature, $\Delta$-temperature, which is the absolute value difference from euthermia at 37$^\circ$ Celsius, turns out to be more useful than body temperature, as the former feature is able to encode both hypothermia and hyperthermia as high values. The model then learns that both conditions are critical, while hyperthermia (and hyperpyrexia) increase sepsis risk proportionally more as indicated by the also high positive weight of the body temperature feature.

Derived features that either put treatment and effect in relation or identify physical regulation issues seem to be especially useful in constructing a predictive sepsis risk score. In principle, deep neural networks are able to construct and exploit such feature combinations automatically, however, our models did not show remarkable differences between linear and deep models (Table \ref{tab:modelcomparison}). We hypothesize that the amount of data is not sufficient to learn richer representations of feature combinations, as the linear models are already very close to the best deep models. The relatively wide 95\% confidence intervals listed in Table \ref{tab:modelcomparison} further support this assumption. 

Integrating pre-existing conditions in the form of 10 most frequent occurring ICD-10 codes helps to strengthen both linear and non-linear models. Relative improvements in performance within a model increase when the time of prediction is moved further away from the onset of sepsis: e.g., in case of the linear model, we observe an AUROC increase of 0.3 points for 4h vs. 2.2 points for 24h before sepsis diagnose. This indicates that pre-existing conditions are useful to identify risk patients shortly after admission. In the hours before sepsis onset the contribution of our CCI ICD-10 features is marginal, however, we see consistent gains over all models and time ranges (Table \ref{tab:modelcomparison}).

Our data-driven approach requires to have sufficient occurrences of each single feature in our data. This is reflected by the constraint that each selected ICD-10 code must be present in at least 20 patients. Due to this decision, we were forced to ignore many sepsis relevant disorders. For example, diabetes is such fine grained in ICD-10 coding that it does not appear in our feature set. We thus explored the idea of grouping similar codes by degrading them to their category, i.e., taking only the first three characters as a feature, but models trained additionally with such category codes did not show any measurable increase in performance. We conclude that the ICD-10 category codes are too coarse to add useful information to the model. For the same reason, a grouping scheme according to \citet{SundararajanETAL:06} is likely to give similar results: the severity and distinction of symptoms within a category or group is relevant information for the outcome. A too coarse grouping scheme will obfuscate this information as it happened in the experiments on ICD-10 categories. Only a larger patient cohort would have enabled us to include more pre-existing conditions. 

A closer look on the learned weights of the 10 fine-grained pre-existing conditions listed in Table \ref{tab:weightsicd} reveals I69.3 (Sequelae of cerebral infarction), K70.3  (Alcoholic cirrhosis of liver), I42.88 (Other cardiomyopathies), and I50.01 (Left ventricular failure) as increasingly high risk conditions for developing sepsis. Surprisingly, two conditions, namely N18.3 (moderate chronic kidney disease) and Z95.5 (coronary angioplasty implant and graft), had negative weights exceeding the standard deviation $\sigma$. This indicates a decrease in severity score if the condition is present in a patient. Besides known thrombocyte inhibition in states of chronic kidney disease \citep{BladelETAL:12}, consulted clinicians identified a common pattern in the treatment for condition Z95.5 (coronary angioplasty implant and graft) as well as for G81.0 (flaccid hemiplegia) and Z95.1 (aortocoronary bypass graft) which also had negative weights: all conditions are commonly treated with acetyl-salicylic acid for reasons of secondary prevention. This medication, however, is known to result in thrombocyte inhibition as well. At the same time, the concentration of thrombocytes is the second highest weighted risk factor in our linear models (Table \ref{tab:weightsfull}). Our findings are consistent with recent meta-analyses on the reduction of mortality risk in patients taking acetyl-salicilic acid prior to sepsis onset \citep{TrauerETAL:17}. 

\begin{table}[t]
\centering
\begin{tabular}{lrrr}
 \toprule
ICD-10 & description & \multicolumn{1}{c}{value} & \multicolumn{1}{c}{$\sigma$} \\
 \midrule
I50.01 & Left ventricular failure & $0.30787$ & $0.0230$ \\
I42.88 & Other cardiomyopathies & $0.26893$ & $0.0336$ \\
K70.3  & Alcoholic cirrhosis of liver & $0.13594$ & $0.0265$ \\
I69.3  & Sequelae of cerebral infarction & $0.13567$ & $0.0213$ \\
J44.99 & Chronic obstructive pulmonary disease, unspecified & $0.00929$ & $0.0234$ \\
C79.3  & 2ry malignant neoplasm of brain and cerebral meninges & $0.00366$ & $0.0193$ \\
 \midrule
N18.3  & Chronic kidney disease, moderate & $-0.07416$ & $0.0398$ \\
Z95.5  & Coronary angioplasty implant and graft & $-0.04883$ & $0.0306$ \\
G81.0  & Flaccid hemiplegia, unspecified & $-0.02385$ & $0.0289$ \\
Z95.1  & Aortocoronary bypass graft & $-0.01421$ & $0.0254$ \\
 \bottomrule
 
\end{tabular}
\caption{Averaged weights and corresponding standard deviations $\sigma$ of the linear model for the 10 most frequent pre-existing conditions by ICD-10 code. Weights are grouped into negatives and positives ordered by absolute value. The presence of pre-conditions with positive weights (upper part) increases the risk score. Pre-conditions with negative weights (lower part) have a decreasing effect on risk score. 
}
\label{tab:weightsicd}
\end{table}

\subsection{Cases of Terminally Ill Patients}

Early antibiotic treatment of patients with sepsis is crucial to decrease risk of progression to more critical stages like severe sepsis and septic shock \citep{WhilesETAL:17,PeltanETAL:17}. Delaying treatment even by a short period of time in the magnitude of hours has shown to increases mortality in several clinical studies \citep{KumarETAL:06, GaieskiETAL:10, FerrerETAL:14}. We thus examined the severity score in detail on six terminally ill patients which died in the ICU or were considered terminally ill and received palliative care at some point. Three patients did not develop sepsis and two of them received palliative care 26 hours and 86 hours after admission. The three remaining patients were considered septic at 170, 192, and 196 hours after ICU admission. All six patients are taken from the same test set and selected such that the duration of monitoring (in case of non-sepsis patients) or the time to sepsis onset is comparable. 

The average age of terminally ill patients is higher than the cohort's, especially in the case of non-sepsis patients (72.7 for non-sepsis, 66.7 for sepsis patients). The terminally ill non-sepsis patients were male patients of age 76, 84, and 58. Among the terminally ill sepsis-patients was one woman aged 85, while the other two males were 47 and 67 years old. In general, this sample size is far too small to draw conclusions. It is still notable that I50.01, a left ventricular failure, was present in the two younger sepsis patients who died after 106 and 20 days in the ICU. This specific ICD-code received the second highest weight among all preconditions in our model, indicating an important influence on the patient's risk of developing sepsis. 

\begin{figure}[t]
  \centering
\begin{minipage}[t]{0.5\linewidth} 
	\includegraphics[angle=-90,width=6cm]{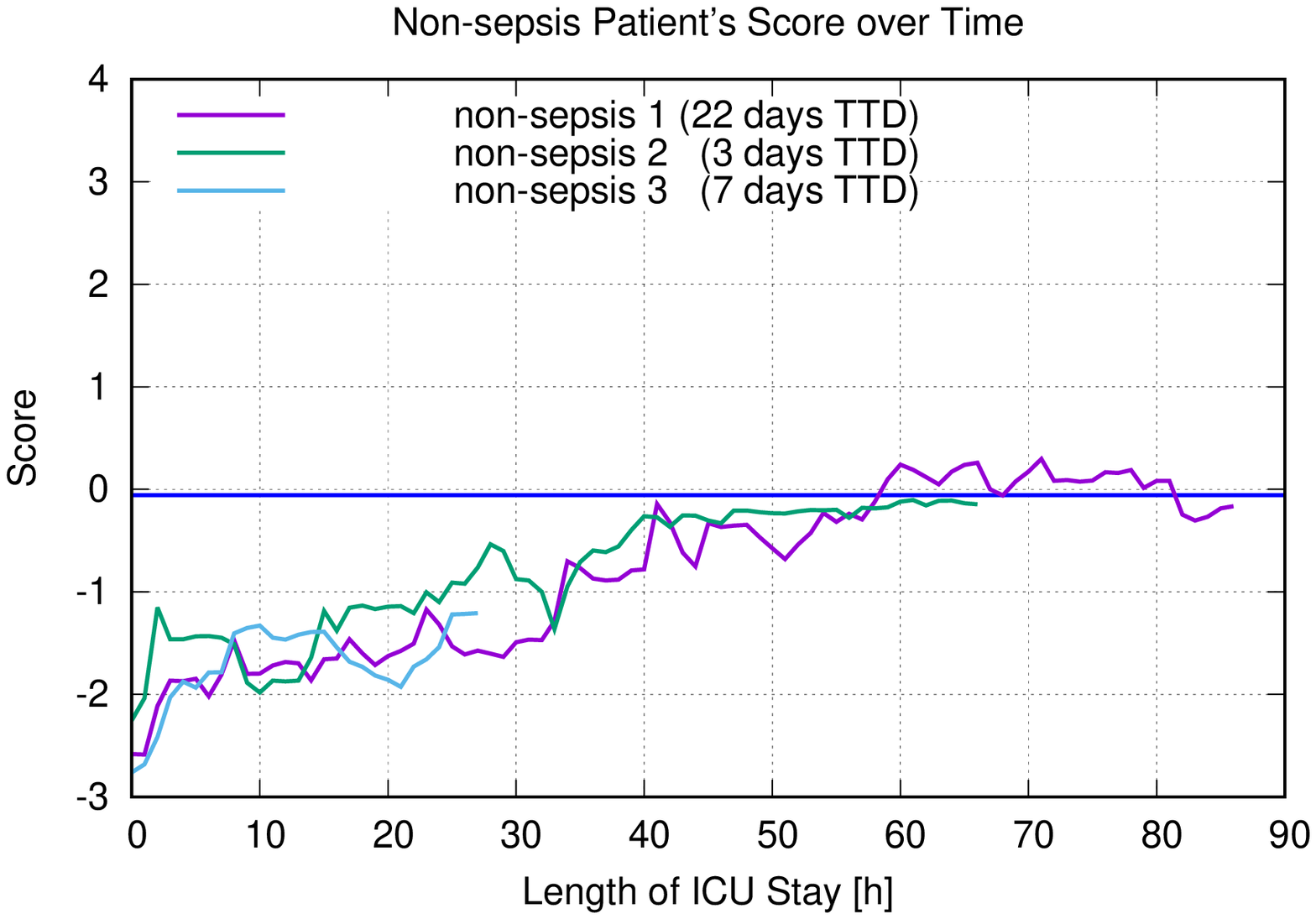}
\end{minipage}
\begin{minipage}[t]{0.49\linewidth} 
	\includegraphics[angle=-90,width=6cm]{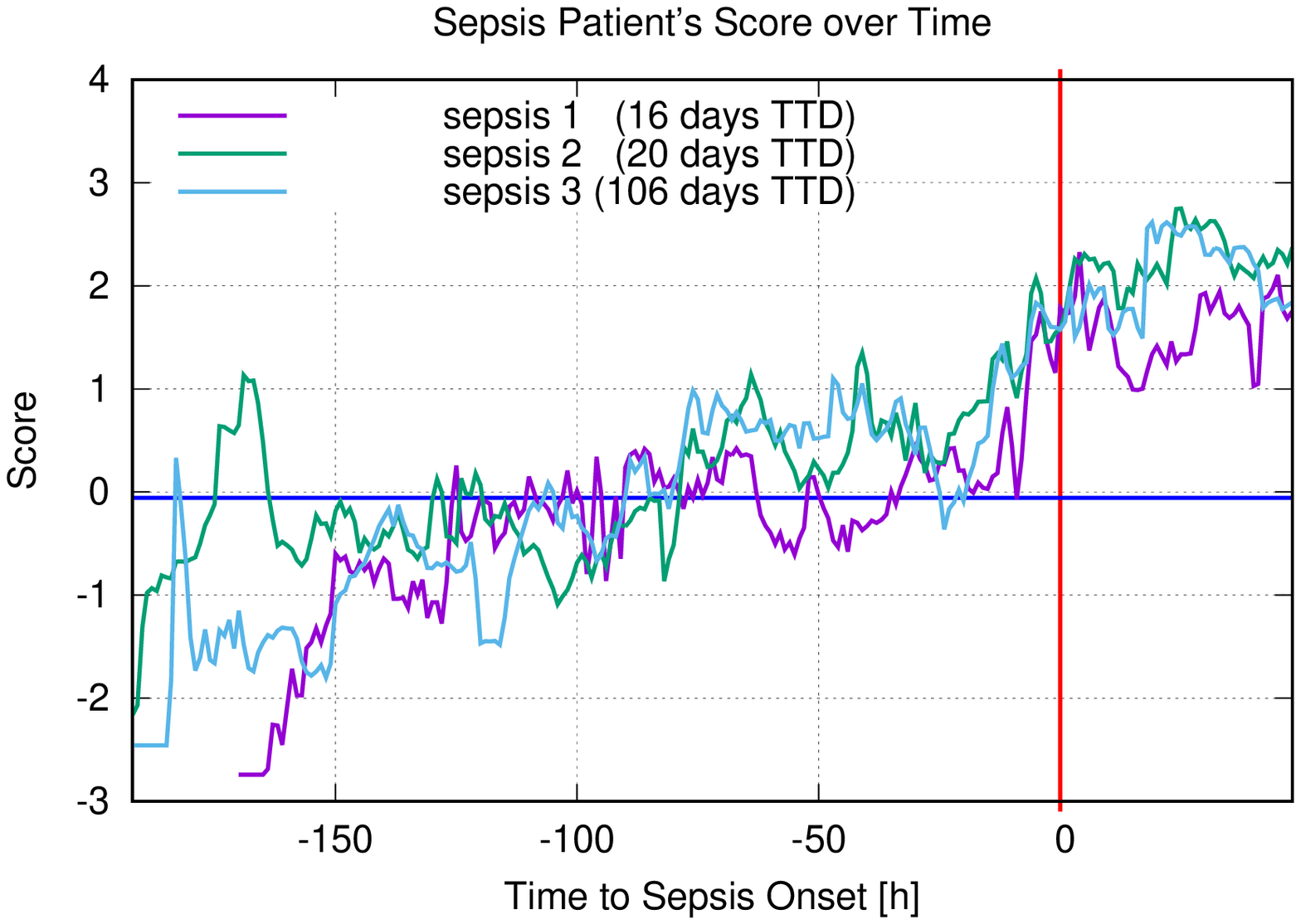}
\end{minipage} 

\caption{Example scores of non-sepsis and sepsis patients who died during their stay or after receiving palliative care (patients ``non-sepsis 1'' and ``non-sepsis 3''). The time-to-death (TTD) in days is calculated from start of admission. The red vertical line in the right figure indicates the onset of the first sepsis episode, the horizontal blue line marks the threshold where the model achieved a sensitivity and specificity of $0.8$ on the test set.}
\label{fig:terminallyill}
\end{figure}

Figure \ref{fig:terminallyill} visualizes the severity score for the sample of three non-sepsis patients (left) and three sepsis patients (right) over time. The evaluated model is the non-linear deep model described in Section \ref{sec:models} with integrated pre-existing conditions. Detailed results are listed in the rightmost column in Table \ref{tab:modelcomparison}. The red vertical line in the right plot indicates the time our experts identified sepsis (at $\text{Time to Sepsis Onset}=0$). The blue horizontal line denotes the threshold at $-0.057$ for which our model achieves a sensitivity of $0.800$ and a specificity of $0.799$. 

Although all six patients were critically ill and died at the end of their stay, the severity score graphs for non-sepsis and sepsis patients are well distinguishable. Even at times preceding the onset of sepsis, the score graph is considerably more often above the threshold line for sepsis patients than for non-sepsis patients. However, simple analysis of the score over time might have predicted sepsis too early: given the indicated threshold, sepsis would have been predicted several days before it was diagnosed by an expert. The model would also predict sepsis for one terminally ill non-sepsis patient. 

Modified sepsis prediction criteria that consider the duration of time and the absolute amount a patient's score exceeds a certain threshold would improve predictive power of the model in case of severely and thus possibly terminally ill patients. If antibiotic treatment was assessed earlier as a result of the severity score, the outcome might have been different for the two younger patients (labeled ``sepsis 2'' and ``sepsis 3'' in Figure \ref{fig:terminallyill}). Several clinical investigations have shown that their chances of survival would have been improved \citep{KumarETAL:06, GaieskiETAL:10, FerrerETAL:14}.

\begin{table}[!t]
\centering
\begin{tabular}{rccc}
 \toprule
feature name [category] & unit & def. value & req. range\\
 \midrule
Respiratory minute volume [\texttt{v}] &  L/min& 7.53 & 0--20 \\
Heart rate [\texttt{v}] & $\text{min}^{-1}$& 75 & 0--400 \\
Temperature [\texttt{v}] & $^\circ$Celsius & 37.0 & 29--42.5 \\
Partial CO$_2$ [\texttt{v}] & mmHg & 40.0 & 10--200 \\
Respiratory rate [\texttt{v}] & $\text{min}^{-1}$& 12.0 & 0--60 \\
Diastolic blood pressure [\texttt{v}] & mmHg & 60.0 & 0--150 \\
Fraction of inspired O$_2$ [\texttt{v}] & \% & 0.3 & 0.21--1 \\
Mean blood pressure [\texttt{v}] & mmHg & 80.0 & 0--180 \\
Systolic blood pressure [\texttt{v}] & mmHg & 120.0 & 0--300 \\
Cardiac output [\texttt{v}] & L/min& 5.0 & 0--40 \\
Oxygen saturation [\texttt{v}] & \% & 99.0 & 0--100 \\
Oxygenation saturation (mixed) [\texttt{v}] & \% & 70.0 & 30--100 \\
Urine output [\texttt{v}] & mL/h & 50 & 0--10000 \\
Net balance [\texttt{v}] & mL/h & 0.0 & -- \\
Stroke volume [\texttt{v}] & mL & 88.0 & 0--250 \\
Systemic Vascular Resistance Index [\texttt{v}] & {$\text{dyn}\cdot\text{s}\cdot\text{cm}^{-5}$} & {2068.0} & {0--8000} \\
Leukocytes [\texttt{l}] & $10^9/\text{L}$& 8.0 & 0--100 \\
Arterial pH [\texttt{l}] & -- & 7.4 & 6.8--7.9 \\
Bilirubin [\texttt{l}] & mg/dL & 0.8 & 0--30 \\
Blood urea nitrogen [\texttt{l}] & mg/dL& 40.0 & 5--500 \\
Creatinine [\texttt{l}] & mg/dL & 1.0 & 0--20 \\
Partial pressure arterial O$_2$ [\texttt{l}] & mmHg & 100.0 & 20--1000 \\
Thrombocytes [\texttt{l}] &  $10^9/\text{L}$& 200.0 & 0--5000 \\
Lactate [\texttt{l}] & mmol/L& 1.0 & 0--30 \\
Bicarbonate [\texttt{l}] & mmol/L& 23.0 & 0--70 \\
C-reactive protein [\texttt{l}] & mg/L& 5.0 & 0--1000 \\
Hemoglobin [\texttt{l}] & g/dL & 12.0 & 1--25 \\
Lymphocytes [\texttt{l}] & $10^3/\mu{}\text{L}$& 2.0 & 0--50 \\
Sodium [\texttt{l}] & mmol/L & 140.0 & 100--200 \\
Pancreatic lipase [\texttt{l}] & Units/L& 80.0 & 0--5000 \\
Procalcitonin [\texttt{l}] & mmol/L & 0.2 & 0--100 \\
Quick score[\texttt{l}]  & INR & 1.0 & 0.5--10 \\
Blood glucose [\texttt{l}] & mg/dL & 100.0 & 0--1000 \\
Base excess [\texttt{l}] &  mmol/L & 0.0 & -30--30 \\
Chloride [\texttt{l}] & mmol/L & 105.0 & 50--150 \\
Calcium (ionized) [\texttt{l}] & mmol/L & 1.25 & 0.25--2 \\
Potassium [\texttt{l}] &  mmol/L & 4.0 & 1--10 \\
Alanine transaminase [\texttt{l}] & Units/L & 40.0 & 0--10000 \\
Aspartate aminotransferase [\texttt{l}] & Units/L& 40.0 & 0--10000 \\
Horowitz index [\texttt{d}] & mmHg & 400 & 20--5000 \\
BUN-to-creatinine ratio [\texttt{d}] & -- & -- & -- \\
$\Delta$-temperature [\texttt{d}] & -- & 0 & -- \\
 \bottomrule

\end{tabular}
\caption{All 42 features extracted from the hospital's EHR system with default values and required ranges. The letter in brackets following the feature name denote the category, i.e. vital and physiological parameters [v], laboratory values [l] and derived features [d].}
\label{tab:stdfeatures}
\end{table}

\section{Conclusion}
\label{sec:conc}
In this paper we presented an approach to machine learning for sepsis prediction that utilizes a ground truth dataset that was annotated for sepsis status by clinical practitioners. The annotation process is based on an electronic questionnaire that is completed on a daily basis by attending physicians judging the sepsis status of patients. An inter-rater agreement study shows that despite implicit factors in the annotation, a very high $\alpha$-reliability of $0.94$ can be achieved. We argue that the common practice of automatically defining ground truth labels for sepsis by applying criteria such as those defined in the SSCG or Sepsis-3 guidelines leads to a circularity if the same criteria are incorporated as features of the machine learning model. Since our ground truth dataset does not directly depend on given guidelines for sepsis definitions, it thus also promotes validity of sepsis research.

We present an experimental evaluation of training and testing linear and non-linear models on this dataset, using 42 standard measurements extracted from EHRs and 10 pre-existing comorbidities as features. Our evaluation shows that linear and non-linear models perform similarly, with best results of $81.7$ AUROC obtained for a non-linear model that combines all available features. We furthermore exploit the intelligibility of the linear model to interpret contributions of features by their learned weights. Firstly, we discuss potential discoveries such as a high thrombocytes level as a risk factor for sepsis. It is unlikely that this finding would result in an approach that defines sepsis labels directly by measuring SOFA scores since high platelets count is defined to yield low SOFA score \citep{VincentETAL:96,SingerETAL:16,SeymourETAL:16}. Furthermore, interpreting learned feature weights allows us to evaluate counterintuitive findings such as a seemingly positive influence of comorbidities on severity of sepsis.

An interesting topic for future work is to go beyond a prediction of sepsis at selected points in time, but to attempt to model the whole time series of clinical measurements for more accurate prediction, and for a potential discovery of new criteria responsible for the development of sepsis.

\section*{Acknowledgements}
This research has been conducted in project SCIDATOS (Scientific Computing for Improved Detection and Therapy of Sepsis), funded by the Klaus Tschira Foundation, Germany (Grant number 00.0277.2015).
  
\bibliography{paper_pre}

\end{document}